\documentclass[10pt]{article}
\usepackage[letterpaper]{geometry}
\usepackage[hidelinks]{hyperref}
\usepackage[utf8]{inputenc}
\usepackage[T1]{fontenc}
\usepackage{amsmath,amsfonts,lmodern,mathtools,nameref,natbib,setspace,tikz,titlesec}
\usetikzlibrary{angles,calc,quotes}

\newcommand{\documentname}{\textsl{Note}}
\sloppypar
\renewcommand{\paragraph}[1]{\par\addvspace{1em}\noindent\textsl{#1}~---}
\newcommand{\secbreak}{\bigskip{\centering\footnotesize%
\rotatebox[origin=c]{55}{$\triangle$}~~~%
\rotatebox[origin=c]{35}{$\triangle$}~~~%
\rotatebox[origin=c]{15}{$\triangle$}\par}\bigskip\noindent}
\titleformat*{\subsection}{\normalfont\normalsize\slshape}
\titlespacing*{\subsection}{0pt}{0pt}{0pt}
\setcitestyle{aysep={}}
\renewcommand{\d}{\mathrm{d}}
\newcommand{\abs}[1]{|\,{#1}\,|}

\usepackage{xcolor}

\newcommand{\aOne}{\textsuperscript{\textasteriskcentered}}
\newcommand{\aTwo}{\textsuperscript{\textdagger}}
\newcommand{\aThree}{\textsuperscript{\textdaggerdbl}}

\newcommand{\aSix}{\textsuperscript{\textparagraph}}

\begin{document}\thispagestyle{empty}\setcounter{secnumdepth}{0}

\section*{\centering\normalsize\uppercase{
A formula for the area of a triangle:\\
Useless, but explicitly in Deep Sets form}}

\medskip
\noindent
\textbf{Connor Hainje}\aOne\aSix{}
\textsl{and}
\textbf{David W. Hogg}\aOne\aTwo\aThree%, and
%\textbf{Soledad Villar}\aFour\aFive

\medskip
{\footnotesize\par\noindent \aOne\textsl{
Center for Cosmology and Particle Physics, Department of Physics, New York University}}
{\footnotesize\par\noindent \aTwo\textsl{
Max-Planck-Insitut f\"ur Astronomie}}
{\footnotesize\par\noindent \aThree\textsl{
Center for Computational Astrophysics, Flatiron Institute}}
%{\footnotesize\par\noindent \aFour\textsl{
%Department of Applied Mathematics and Statistics, Johns Hopkins University}}
%{\footnotesize\par\noindent \aFive\textsl{
%Mathematical Institute for Data Science, Johns Hopkins University}}
{\footnotesize\par\noindent \aSix\textsl{
Email:} \texttt{\href{mailto:connor.hainje@nyu.edu}{connor.hainje@nyu.edu}}}
{\footnotesize\par}  % needed to prevent some weird spacing issue

\smallskip
\paragraph{Abstract}
Any permutation-invariant function of data points $\vec{r}_i$
can be written in the form $\rho(\sum_i\phi(\vec{r}_i))$ for suitable functions $\rho$ and $\phi$.
This form---known in the machine-learning literature as Deep Sets---also generates a map--reduce algorithm.
The area of a triangle is a permutation-invariant function of the locations $\vec{r}_i$ of the three corners $1\leq i\leq 3$.
We find the polynomial formula for the area of a triangle that is explicitly in Deep Sets form.
This project was motivated by questions about the fundamental computational complexity of $n$-point statistics in cosmology;
that said, no insights of any kind were gained from these results.

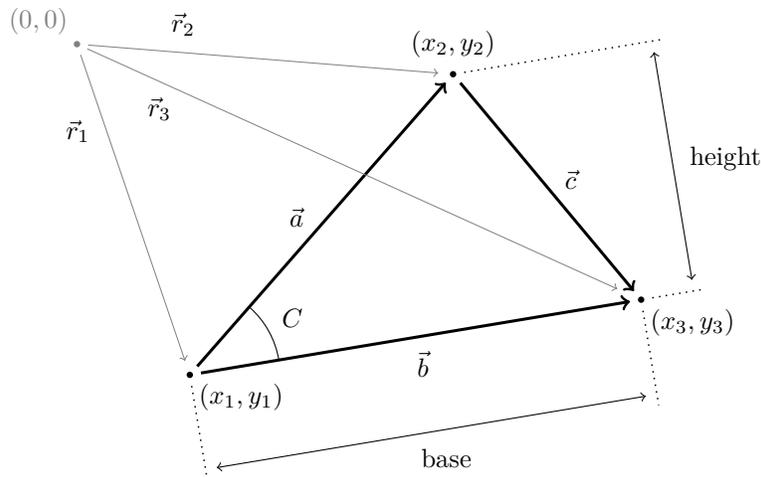
\begin{figure}[t!]
    \centering
    \begin{tikzpicture}
    \coordinate (F) at (-1.5,4.4);
    \coordinate (O) at (0,0);
    \coordinate (A) at (3.5,4);
    \coordinate (B) at (6,1);

    \pgfmathsetmacro{\a}{sqrt(3.5*3.5 + 4*4)}
    \pgfmathsetmacro{\b}{sqrt(6*6 + 1*1)}
    \pgfmathsetmacro{\c}{sqrt((6-3.5)*(6-3.5) + (4-1)*(4-1))}
    \pgfmathsetmacro{\s}{(\a+\b+\c)/2}
    \pgfmathsetmacro{\h}{2*sqrt(\s*(\s-\a)*(\s-\b)*(\s-\c))/\b}

    \pgfmathsetmacro{\DeltaX}{6 - 0}  % Bx - Ox
    \pgfmathsetmacro{\DeltaY}{1 - 0}  % By - Oy
    \pgfmathsetmacro{\len}{sqrt(\DeltaX*\DeltaX + \DeltaY*\DeltaY)}

    \def\bLen{0.5mm}
    \def\hLen{0.3mm}
    \pgfmathsetmacro{\bdx}{-\DeltaY/\len * \bLen}
    \pgfmathsetmacro{\bdy}{\DeltaX/\len * \bLen}
    \pgfmathsetmacro{\hdx}{\DeltaX/\len * \hLen}
    \pgfmathsetmacro{\hdy}{\DeltaY/\len * \hLen}
    \pgfmathsetmacro{\hx}{-\DeltaY/\len * \h}
    \pgfmathsetmacro{\hy}{\DeltaX/\len * \h}

    \coordinate (BaseO) at ($(O) - (\bdx,\bdy)$);
    \coordinate (BaseB) at ($(B) - (\bdx,\bdy)$);
    \coordinate (HeightO) at ($(B) + (\hdx,\hdy)$);
    \coordinate (HeightA) at ($(B) + (\hdx,\hdy) + (\hx,\hy)$);

    % mark points
    \fill (O) circle (1.2pt) node[below right]{$(x_1, y_1)$};
    \fill (A) circle (1.2pt) node[above, yshift=+1mm]{$(x_2, y_2)$};
    \fill (B) circle (1.2pt) node[below right]{$(x_3, y_3)$};
    \fill[gray] (F) circle (1.2pt) node[above left]{$(0, 0)$};

    % draw vectors
    \def\pad{1.5mm}
    \draw[->, line width=0.4mm]
        ($(O)!\pad!(A)$) -- ($(A)!\pad!(O)$)
        node[midway, left, xshift=-1mm, yshift=+1mm] {$\vec{a}$};
    \draw[->, line width=0.4mm]
        ($(O)!\pad!(B)$) -- ($(B)!\pad!(O)$)
        node[midway, below, xshift=+1mm, yshift=-0.5mm] {$\vec{b}$};
    \draw[->, line width=0.4mm]
        ($(A)!\pad!(B)$) -- ($(B)!\pad!(A)$)
        node[midway, right, xshift=+1mm, yshift=+1mm] {$\vec{c}$};

    \draw[->, line width=0.1mm, gray]
        ($(F)!\pad!(O)$) -- ($(O)!2mm!(F)$)
        node[near start, left, xshift=-1mm, yshift=0mm, black] {$\vec{r}_1$};
    \draw[->, line width=0.1mm, gray]
        ($(F)!\pad!(A)$) -- ($(A)!2mm!(F)$)
        node[near start, above, xshift=+1mm, yshift=+1mm, black] {$\vec{r}_2$};
    \draw[->, line width=0.1mm, gray]
        ($(F)!\pad!(B)$) -- ($(B)!3.5mm!(F)$)
        node[very near start, right, xshift=-2mm, yshift=-4mm, black] {$\vec{r}_3$};

    % draw arc for angle
    \draw pic["$C$", draw=black, angle radius=12mm, angle eccentricity=1.3] {angle=B--O--A};

    % draw base, height guide lines
    \draw[dotted, line width=0.2mm]($(O)!\pad!(BaseO)$) -- ($(BaseO)$);
    \draw[dotted, line width=0.2mm]($(B)!\pad!(BaseB)$) -- ($(BaseB)$);
    \draw[dotted, line width=0.2mm]($(B)!\pad!(HeightO)$) -- ($(HeightO)$);
    \draw[dotted, line width=0.2mm]($(A)!\pad!(HeightA)$) -- ($(HeightA)$);

    % draw base, height size lines
    \draw[<->]
        ($(BaseO)!\pad!(O)!\pad!(BaseB)$) -- ($(BaseB)!\pad!(B)!\pad!(BaseO)$)
        node[midway, below, xshift=+2mm, yshift=-1mm] {base};
    \draw[<->]
        ($(HeightO)!\pad!(O)!\pad!(HeightA)$) -- ($(HeightA)!\pad!(A)!\pad!(HeightO)$)
        node[midway, right, xshift=+1mm, yshift=+1mm] {height};
    
    \end{tikzpicture}
    \caption{A triangle.}
    \label{fig:triangle}
\end{figure}

\secbreak
The area $\Delta$ of a triangle can be written in many ways.
Here are a few:
\begin{align}
\Delta 
&= \frac{1}{2} \, (\text{base}) \, (\text{height}) \label{eq:school} \\
&= \frac{1}{2}\,a\,b\,\sin(C) \label{eq:sine} \\
&= \frac{1}{2}\, \abs{\vec{a} \times \vec{b}}
    = \frac{1}{2}\, \abs{\vec{a} \wedge \vec{b}}\label{eq:cross} \\
&= \frac{1}{2}\, \abs{
    x_1 \, y_2 - x_2 \, y_1 +
    x_2 \, y_3 - x_3 \, y_2 +
    x_3 \, y_1 - x_1 \, y_3
}\label{eq:polynomial} \\
&= \sqrt{s\,(s-a)\,(s-b)\,(s-c)} \label{eq:Heron} ~,
\end{align}
where the quantities used are defined in Figure~\ref{fig:triangle}, apart from the side lengths
$a, b, c = \abs{\vec{a}}, \abs{\vec{b}}, \abs{\vec{c}}$
and the semi-perimeter $s = (a + b + c)/2$.
Formula~\eqref{eq:school} is the usual introduction to the subject, traceable back to \citet{Euclid300BC}.
Formula~\eqref{eq:sine} makes use of the fact that if the base is $b$ then the height is $a\,\sin(C)$.
Formula~\eqref{eq:cross} generalizes to triangles in higher dimensional spaces, using the fact that the cross or wedge product has a magnitude equal to the area of the parallelogram spanned by $\vec{a}$ and $\vec{b}$.
Formula~\eqref{eq:polynomial} gives the absolute value of a polynomial in terms of the Cartesian coordinates of the corners in two dimensions; it is equal to an expansion of \eqref{eq:cross}.
Formula~\eqref{eq:Heron} is Heron's formula, which elegantly computes the area from the side lengths alone.
For a collection of 110 unique triangle area formulae, see \citet{Baker1885a,Baker1885b}.

The area of a triangle obeys many symmetries. It is invariant to translation, rotation, and reflection, as well as to the labeling of sides or corners. However, these symmetries are not always obvious from a given formula; for example, of those listed above, only the absolute-value polynomial \eqref{eq:polynomial} and Heron's formula \eqref{eq:Heron} are explicitly permutation-invariant.

A function $f(\vec{r}_1, \dots, \vec{r}_N)$ is invariant to permutation of its arguments if and only if it can be decomposed in the form
\begin{equation}
    f(\vec{r}_1, \dots, \vec{r}_N) = \rho \, \big( \sum_{i=1}^{N} \, \phi(\vec{r}_i) \big)~,\label{eq:DeepSets}
\end{equation}
for some functions $\rho$ and $\phi$, where $\phi$ is usually vector valued.
This result---often known by the name of the deep learning architecture derived from it, \emph{Deep Sets} (\citealt{Zaheer+17deepsets}; see Theorem~2)---is not at all obvious.
For example, even though Heron's formula \eqref{eq:Heron} is explicitly permutation-invariant, it is not actually in the form of \eqref{eq:DeepSets}.\footnote{But see equation \eqref{eq:HeronDeepSets} below.}
Further, despite the triangle area being permutation-invariant, there is no known (prior to this work) formula for the area of a triangle that is explicitly in Deep Sets form.

While this result was recently popularized as Deep Sets, it follows from the classical theory of invariant polynomials.
Any permutation-invariant (or symmetric) polynomial of scalar data $f(\{r_i\})$ can be written as a polynomial of the power sums $\sum_i r_i^m$ \citep{Waring1782}.
Further, symmetric polynomials of vector data $\{ \vec{r}_i \}$ are decomposable into polarizations of the scalar symmetric polynomials \citep{Weyl1939}.
Together, these imply that any permutation-invariant polynomial of vector data can be written as a polynomial in the polarizations of the power sums, which look like sums of the form $\sum_i \, (\vec{r}_{i})_{j}^{m_j} \cdots (\vec{r}_{i})_{k}^{m_k}$, where $(\vec{r}_i)_j$ is the $j^{\rm th}$ component of $\vec{r}_i$.
It also matches onto the Deep Sets form, where the polarizations of the power sums are the components of the vector $\sum_i \phi(\vec{r}_i)$, which are combined in a polynomial $\rho$.

Although a Deep Sets formula for the area of a triangle is of minimal, or even zero, practical interest, our search is motivated by real problems in physics.
The laws of classical physics are generally invariant to permutation.
For example, the gravitational (or electric) potential energy of a set of point masses (or charges) is invariant to the ordering or labeling of the points. 
Specifically, for a system of $N$ point particles with masses $m_i$ at positions $\vec{r}_i$, the gravitational potential energy is
\begin{equation}
\label{eq:EnergyNaive}
    U = -G \, \sum_{i=1}^{N} \sum_{j=i+1}^{N} \frac{m_i \, m_j}{\abs{\vec{r}_i - \vec{r}_j}} ~.
\end{equation}
This is explicitly permutation invariant since it involves a sum without repeats over all pairs of particles, but it is not in Deep Sets form \eqref{eq:DeepSets} because it involves a double sum---with $\mathcal{O}(N^2)$ computational scaling---of quantities computed from pairs of particles.

Further, and important to our own motivations, cosmological clustering statistics such as 
the correlation function \citep{Peebles1973},
the power spectrum \citep{Peebles1973},
the three-point correlation function \citep{PeeblesGroth1975},
the bispectrum \citep{FrySeldner1982},
and higher-order statistics \citep{Peebles1980book}
used for precision measurement in cosmology \citep[e.g.,][]{Eisenstein+2005BAO,Planck18PNG,Planck18Inflation,Cabass+2022}
are all permutation-invariant.
Calculating these statistics na\"ively involves loops over all $k$-tuplets of points, which is not only inconsistent with Deep Sets form, but also has a runtime complexity of $\mathcal{O}(N^k)$, where $N$ is the number of points in the set.

Because all these (computationally expensive) statistics are permutation-invariant, it must be possible to write them in Deep Sets form.
At the same time, Deep Sets form \eqref{eq:DeepSets} looks like it is computable in linear $\mathcal{O}(N)$ time.
What resolves this discrepancy?
The resolution must be either that $\phi$ has a dimensionality that effectively grows with $N$, or else that the complexity of $\rho$ effectively grows with $N$.
Still, these time complexities may be better than the $\mathcal{O}(N^k)$ runtime of na\"ive implementations.

Even if $\phi$ and $\rho$ grow rapidly with $N$, it might be that Deep Sets form \eqref{eq:DeepSets} delivers a computational advantage in that it is explicitly in map--reduce form.
Map--reduce is a model for parallelizing and distributing a calculation over a large dataset \citep{DeanGhemawat2008,Lammel2008}.
In our case, $\phi$ is the ``map'':
It can be executed independently and asynchronously on each data point.
The sums are then the ``reduce,'' which can be performed hierarchically across a data center such that the wall-clock runtime of the sums scales only logarithmically $\mathcal{O}(\log N)$ with the number of points.
(Granted, a triangle will never have more than three points!)

So, there may be benefits to finding Deep Sets forms for problems that depend on permutation-invariant pairs, triplets, or $k$-tuplets of points.
The simplest such problem is the area of a triangle, 
which depends on pairs of points via
side lengths, as in Heron's formula \eqref{eq:Heron};
angles, as in \eqref{eq:sine};
point displacements, as in \eqref{eq:cross};
or as a sum over $x_i \, y_j - x_j \, y_i$, as in \eqref{eq:polynomial}.

\secbreak
We find this formula---in Deep Sets form---for the area $\Delta$ of a triangle:
\begin{align}
    \Delta^2 = \ &
    \frac{3}{4}
        \,\big( \sum_{i=1}^{3} x_i^2 \big)
        \,\big( \sum_{i=1}^{3} y_i^2 \big)
    - \frac{3}{4}
        \,\big( \sum_{i=1}^{3} x_i \, y_i \big)^2 
    + \frac{2}{4}
        \,\big( \sum_{i=1}^{3} x_i \, y_i \big)
        \,\big( \sum_{i=1}^{3} x_i \big)
        \,\big( \sum_{i=1}^{3} y_i \big)
    \nonumber\\ &
    - \frac{1}{4}
        \,\big( \sum_{i=1}^{3} x_i^2 \big)
        \,\big( \sum_{i=1}^{3} y_i \big)^2
    - \frac{1}{4}
        \,\big( \sum_{i=1}^{3} y_i^2 \big)
        \,\big( \sum_{i=1}^{3} x_i \big)^2
     ~.
\label{eq:result}
\end{align}
This formula \eqref{eq:result} for the squared area\footnote{%
    We choose to work with squared area $\Delta^2$ and not area $\Delta$ because, geometrically, areas are signed quantities.
    Indeed, the formulae \eqref{eq:school} through \eqref{eq:Heron} either (implicitly) involve square roots of polynomials, or else absolute value signs, such that the corresponding pure polynomial expression would be for the squared area.}
is the primary contribution of this \documentname, along with the generalizations below.

This equation \eqref{eq:result} is verified via computational algebra, expanding and simplifying the formula to demonstrate its equivalence to \eqref{eq:polynomial}.
The formula is in Deep Sets form, since
every expression in parentheses in \eqref{eq:result} is a pure sum over all three points.
Being very explicit, it matches \eqref{eq:DeepSets} for $\rho$ and $\phi$ defined as follows:
\begin{gather}
    \label{eq:resultDeepSets}
    \phi(\vec{r}_i) = \big[
        x_i, \,
        y_i, \,
        x_i^2, \,
        x_i \, y_i, \,
        y_i^2
    \big], \quad
    \big[
        F_{10}, \,
        F_{01}, \,
        F_{20}, \,
        F_{11}, \,
        F_{02}
    \big] = \sum_{i=1}^{3} \phi(\vec{r}_i),
    \\
    \rho \, \big( \sum_{i=1}^{3} \phi(\vec{r}_i) \big)
    = \frac{3}{4} \, F_{20} \, F_{02}
    - \frac{3}{4} \, F_{11}^2
    + \frac{2}{4} \, F_{11} \, F_{10} \, F_{01}
    - \frac{1}{4} \, F_{20} \, F_{01}^2
    - \frac{1}{4} \, F_{02} \, F_{10}^2~.
    \nonumber
\end{gather}
Notice that the runtime is formally linear in the number of points, since $\phi$ and $\rho$ can be considered to have constant $\mathcal{O}(1)$ time complexity.
However, because the dimensionality of $\phi$ is larger than the number of points, evaluating the triangle area in this way will probably not ever be faster than a loop over pairs of points.

The area of a triangle is not only invariant to permutation, but also to translation, rotation, and reflection.
Formulae \eqref{eq:sine} and \eqref{eq:Heron}, by virtue of referencing only side-lengths and angles, are explicitly invariant to these other symmetries.
While true, it is not obvious that our solution \eqref{eq:result} is so invariant.
An explicit proof of this is left to the reader.

Recall that a triangle with vertices $\{ \vec{r}_1, \vec{r}_2, \vec{r}_3 \}$ has one-half the area of the parallelogram spanned by the displacement vectors $\{ \vec{r}_2 - \vec{r}_1, \vec{r}_3 - \vec{r}_1 \}$.
So our result \eqref{eq:result} gives an integer polynomial for the squared area of this parallelogram.
Using this fact, we have written code that performs a symbolic regression to recover our result.\footnote{
    The simple code is available for re-use under an open-source license at \url{https://github.com/davidwhogg/TriangleAreas}.} 
The code performs least-squares regression to find a fourth-order homogeneous polynomial of the multi-power sums $\sum_i \, x_i^m \, y_i^n$, and it requires that the result gives the squared areas to machine precision after rounding the coefficients to the nearest integers.
Using and extending the code, we also find several equivalent but more complicated formulae, which are given in the \nameref{sec:appendix}.
All the additional formulae require a higher-dimensional $\phi$ with higher-order terms like $x_i^2 \, y_i^2$ which are combined in ways that reduce to our result \eqref{eq:result}, indicating that \eqref{eq:result} may be the ``simplest'' such formula.

If the triangle is embedded in a higher-dimensional space with vertex positions given by $\vec{r}_i \in \mathbb{R}^{d}$ for $d \geq 2$, the area of the triangle can be determined by summing in quadrature the areas of the triangles made by all pairs of mutually orthogonal projections of the points:
\begin{equation}
    \Delta^2 = \sum_{1 \leq j < k \leq d} \Delta_{jk}^2~, 
    \label{eq:quadrature}
\end{equation}
where $\Delta_{jk}$ is the area of the two-dimensional triangle formed by just the $j^{\rm th}$ and $k^{\rm th}$ components of the position vectors.
A proof is given in the \nameref{sec:appendix}.
By using \eqref{eq:result} for each of the two-dimensional sub-areas, each can be computed in a permutation-invariant way and the entire result is thus explicitly permutation invariant and in Deep Sets form.

The result can also be extended to the volume of a $d$-dimensional simplex.
A $d$-simplex is defined by $d+1$ vectors $\{ \vec{r}_1, \dots, \vec{r}_{d+1} \}$ in $\mathbb{R}^d$.
Then, the volume $V$ of the $d$-simplex is given by
\begin{equation}
    V^2 = \frac{1}{(d!)^2} \, \det \mathsf{M}~,
    \quad
    \mathsf{M} \equiv \left[\begin{array}{cccc}
        \sum_i \, r_{i1}^2
            & \cdots 
            & \sum_i \, r_{i1} \, r_{id}
            & \sum_i \, r_{i1} \\
        \vdots
            & \ddots 
            & \vdots
            & \vdots \\
        \sum_i \, r_{id} \, r_{i1}
            & \cdots 
            & \sum_i \, r_{id}^2
            & \sum_i \, r_{id} \\
        \sum_i \, r_{i1}
            & \cdots 
            & \sum_i \, r_{id}
            & d + 1 \\
    \end{array}\right]~,
    \label{eq:dsimplex}
\end{equation}
where $r_{ij}$ is the $j^{\rm th}$ component of $\vec{r}_i$.\footnote{%
    This generalization was pointed out by Kendrick Smith (Perimeter).}
A proof is deferred to the \nameref{sec:appendix}.
This is in Deep Sets form, as the components of $\mathsf{M}$ are all of the form $\sum_i \phi(\vec{r}_i)$.

A few notes we'd make about this formula are as follows.
This formula with $d=2$ gives back exactly our result \eqref{eq:result}, since the 2-simplex is a triangle and the 2-volume is its area.
The components of $\mathsf{M}$ are in fact the polarizations of the first and second power sums, so the determinant of $\mathsf{M}$ gives the polynomial in these power sums that we were guaranteed by invariant theory.
The runtime complexity can again be considered to be linear, with $\rho$ and $\phi$ having constant complexity.
However, the dimensionality of $\phi$ is $(d+1)(d+2)/2$, the number of unique components in $\mathsf{M}$, so there are no real performance improvements.
As with the relationship between triangles and parallelograms, the volume of the $d$-parallelotope spanned by the displacement vectors $\{ \vec{r}_2 - \vec{r}_1, \dots, \vec{r}_{d+1} - \vec{r}_1 \}$ is $d!$ times the volume of a $d$-simplex; thus \eqref{eq:dsimplex} is an integer polynomial for the squared volume of this $d$-parallelotope.

\secbreak
While the triangle area, by virtue of being a sum over pairs of points, is in a form similar to some physics problems, the polynomial methods we employ are not so relevant for those problems.
In physics contexts it is common to encode a set of points by defining a density field that is a sum of delta functions located at the points.
Notice that the Fourier transform of this field at a wavevector $\vec{k}$ is
\begin{equation}
    \mathcal{D}_{\vec{k}} = \mathcal{F} \,\bigg\{
        \sum_{j=1}^{N} \delta(\vec{r} - \vec{r}_j)
    \bigg\}
    = \sum_{j=1}^{N} e^{-i \, \vec{k} \cdot \vec{r}_j}~,
\end{equation}
which is exactly in the form of $\sum_i \phi(\vec{r}_i)$.
Hence, any function of the Fourier transform $\mathcal{D}$ of the density field is in Deep Sets form and has permutation symmetry guaranteed.
This is physically useful also because the Fourier basis $\exp (i\, \vec{k}\cdot\vec{x})$ is the eigenbasis of the translation operator, so this choice of the map $\phi$ naturally encodes translation symmetry.\footnote{%
    Interestingly, a transformation like $\phi(r) = \exp(i\,k\,r)$ is often used in natural language processing.
    Most large language models, such as GPT-3 \citep{Brown2020GPT3}, are transformer models \citep{Vaswani2017Attention}, which are fundamentally permutation invariant.
    In order to use the positional information of tokens (which are like words), the tokens are augmented by a ``positional encoding.''
    One of the most popular choices for this encoding is the Rotary Positional Encoding \citep[RoPE;][]{Su2024Roformer}, which takes the word embedding vector for the token in position $m$ and multiplies its components by $e^{i \, m \, \theta}$ for preset constant values $\theta \in \mathbb{R}$.
    While this actually \emph{breaks} permutation invariance rather than enforcing it (since the encoding explicitly involves the index $m$), it underscores the broad applicability of these kinds of transformations.}

Because any function of the Fourier transform $\mathcal{D}$ of the density field is automatically in Deep Sets form,
we can find a Deep Sets form for the gravitational potential energy of a system of point masses.
To be explicit: Consider a system of $N$ point particles with masses $m_i$ and positions $\vec{r}_i$.
The total energy is given as a function of $\mathcal{D}$, and thus in Deep Sets form, by
\begin{align}
    &\phi_{\vec{k}}(m_j, \vec{r}_j)
    =m_j \, e^{-i \, \vec{k} \cdot \vec{r}_j}
    ~,
    \quad
    \sum_{i=1}^{N} \phi_{\vec{k}}(m_i, \vec{r}_i)
    = \mathcal{D}_{\vec{k}}~,
    \nonumber\\
    U = \ &\rho \, \big( \sum_{i=1}^{N} \, \phi_{\vec{k}}(m_i, \vec{r}_i) \big)
    = -\frac{G}{4 \pi^2} \,
        \int \d^3 r \,
        \mathcal{F}^{-1}\!\left\{
            \mathcal{D}_{\vec{k}} \ast \left(
            k^{-2} \, \mathcal{D}_{\vec{k}} \right)
        \right\}~.
    \label{eq:EnergyDeepSets}
\end{align}
A derivation is given in the \nameref{sec:appendix}.
In principle this requires $\phi$ to be evaluated at every possible wavevector $\vec{k}$.
To avoid making this infinite-dimensional $\phi$, one can instead evaluate $\mathcal{D}_{\vec{k}}$ at a finite set of wavevectors $\vec{k}$ and live with approximations to the integrals necessary for the convolution and inverse Fourier transform.
The approximation will improve as the size of $\phi$ grows.

Approaches akin to this Fourier approach can be used \citep[and are used; e.g.,][]{SlepianEisenstein2016,Portillo+2018,Philcox+2022ENCORE} to estimate or approximate $k$-point statistics in cosmology, which was our motivating question.
However, these approaches are not in Deep Sets form, as they generally put the points on a grid and use the Fast Fourier Transform to compute a multipole expansion in the spherical harmonics.
This has linearithmic $\mathcal{O}(N_g \log N_g)$ runtime complexity in the number of grid cells $N_g$, since the Fast Fourier Transform is not in map--reduce form.
Perhaps there are gains to be made by considering the problem explicitly in terms of $\mathcal{D}$ for a small set of wavevectors, which could be computed in map--reduce form.

Of course, these Fourier methods can be used to find another triangle area formula.
\begin{align}
    \phi_{\vec{k}} (\vec{r}_j) &= e^{-i \, \vec{k} \cdot \vec{r}_j}~,
    \quad
    \sum_{j=1}^{3} \phi_{\vec{k}}(\vec{r}_j) = \mathcal{D}_{\vec{k}}~,
    \quad
    \mathcal{P}(\vec{r}) = (2\pi)^d \, \mathcal{F}^{-1}\!\left\{
    \abs{ \mathcal{D}_{\vec{k}} }^2 \right\}~,
    \nonumber\\
    P_m &= \int_{\mathbb{R}^d} \d^d r \, \mathcal{P}(\vec{r}) \, \frac{\abs{\vec{r}}^m}{2}
    = a^m+b^m+c^m
    ~,
    \quad
    \Delta
    = \frac{1}{4} \, \sqrt{P_2^2 - 2 \, P_4}~.
    \label{eq:Fresult}
\end{align}
A complete derivation is provided in the \nameref{sec:appendix}.
Briefly, $\mathcal{P}$ is the Patterson function\footnote{%
    It is also the two-point correlation function $\xi(\vec{r})$ without isotropy, but our usage here is more directly inspired by the crystallography method.}
used in x-ray crystallography \citep{Patterson1934} which has six delta function peaks corresponding to the three displacement vectors and their negatives.
Multiplying the Patterson function by a function of the magnitude $\abs{\vec{r}}$ and integrating over all space gives a sum of the values of the function evaluated at the three side lengths. 
Choosing this function properly allows us to write the $m^{\rm th}$ power sum $P_m$ of the side lengths in this form.

To verify this formula \eqref{eq:Fresult}, note that we can write it in terms of the side lengths as
\begin{equation}
\label{eq:HeronDeepSets}
    \Delta = \frac{1}{4} \, \sqrt{
        P_2^2
        - 2 \, P_4
    }
    = \frac{1}{4} \, \sqrt{
        \left( a^2 + b^2 + c^2 \right)^2
        - 2 \, \left( a^4 + b^4 + c^4 \right)
    }~,
\end{equation}
which is equivalent to Heron's ancient formula \eqref{eq:Heron}.
It is even in Deep Sets form as a function of the side lengths.

\secbreak
In this \documentname, we were motivated by the question of how to convert $k$-point functions in cosmology, which involve loops over $k$-tuplets of points and na\"ively scale as $N^k$, into Deep Sets forms \eqref{eq:DeepSets}, which na\"ively scale as $N$, though perhaps with very large constant-runtime computational costs.
These conversions must be possible, since all permutation-invariant functions can be written in Deep Sets form.
The simplest permutation-invariant problem involving a loop over pairs of points is the computation of the area of a triangle, for which no traditional formula is in Deep Sets form.
We deliver two new formulae, a polynomial \eqref{eq:result} and a Fourier expression \eqref{eq:Fresult}.
These triangle formulae have no practical value, and we also do not solve any question of practical importance for cosmology.

{\footnotesize\par\bigskip\noindent
This project would not have been possible without originating inspiration and discussions with Soledad Villar (JHU).
The authors give special thanks to Ben Blum-Smith (JHU) and Kendrick Smith (Perimeter) for insightful and important input.
The authors also thank
  % Ben Blum-Smith (JHU),
  Jonathan Goodman (NYU),
  Wilson Gregory (JHU),
  Teresa Huang (Flatiron),
  Alexander Novara (NYU),
  Mike O'Neil (NYU),
  Simon Park (Princeton),
  Oliver Philcox (Columbia),
  % Kendrick Smith (Perimeter),
  Kate Storey-Fisher (Stanford),
  Ben Wandelt (JHU),
  Kaze Wong (JHU),
and the Blanton--Hogg group meeting at NYU for valuable comments.
The Flatiron Institute is a division of the Simons Foundation.
CH is supported by the National Science Foundation Graduate Research Fellowship under Grant Number DGE-2234660.
The authors also note that, while we believe the formulae are (barely) novel, it is difficult to be sure, given the challenge of conducting a complete literature review for this ancient topic.\par}

\secbreak
% need a line break this time

\vspace{-\bigskipamount}
\renewcommand{\section}[2]{}%
{\small\singlespacing\bibliography{main}\par}

\begin{thebibliography}{}
\expandafter\ifx\csname natexlab\endcsname\relax\def\natexlab#1{#1}\fi
\providecommand{\url}[1]{\href{#1}{#1}}
\providecommand{\dodoi}[1]{doi:~\href{http://doi.org/#1}{\nolinkurl{#1}}}
\providecommand{\doeprint}[1]{\href{http://ascl.net/#1}{\nolinkurl{http://ascl.net/#1}}}
\providecommand{\doarXiv}[1]{\href{https://arxiv.org/abs/#1}{\nolinkurl{https://arxiv.org/abs/#1}}}

\bibitem[{Baker(1885{\natexlab{a}})}]{Baker1885a}
Baker, M. 1885{\natexlab{a}}, A Collection of Formulae for the Area of a Plane
  Triangle, {\em Annals of Mathematics}, 1, 134

\bibitem[{Baker(1885{\natexlab{b}})}]{Baker1885b}
Baker, M. 1885{\natexlab{b}}, A Collection of Formulae for the Area of a Plane
  Triangle, {\em Annals of Mathematics}, 2, 11

\bibitem[{Brown {et~al.}(2020)Brown, Mann, Ryder, Subbiah, Kaplan, Dhariwal,
  Neelakantan, Shyam, Sastry, Askell, {et~al.}}]{Brown2020GPT3}
Brown, T., Mann, B., Ryder, N., {et~al.} 2020, Language models are few-shot
  learners, {\em Advances in Neural Information Processing Systems}, 33, 1877

\bibitem[{{Cabass} {et~al.}(2022){Cabass}, {Ivanov}, {Philcox},
  {Simonovi{\'c}}, \& {Zaldarriaga}}]{Cabass+2022}
{Cabass}, G., {Ivanov}, M.~M., {Philcox}, O. H.~E., {Simonovi{\'c}}, M., \&
  {Zaldarriaga}, M. 2022, {Constraints on multifield inflation from the BOSS
  galaxy survey}, {\em Physical Review D}, 106, 043506

\bibitem[{Dean \& Ghemawat(2008)}]{DeanGhemawat2008}
Dean, J., \& Ghemawat, S. 2008, MapReduce: simplified data processing on large
  clusters, {\em Communications of the ACM}, 51, 107

\bibitem[{{Eisenstein} {et~al.}(2005){Eisenstein}, {Zehavi}, {Hogg},
  {Scoccimarro}, {Blanton}, {Nichol}, {Scranton}, {Seo}, {Tegmark}, {Zheng},
  {Anderson}, {Annis}, {Bahcall}, {Brinkmann}, {Burles}, {Castander},
  {Connolly}, {Csabai}, {Doi}, {Fukugita}, {Frieman}, {Glazebrook}, {Gunn},
  {Hendry}, {Hennessy}, {Ivezi{\'c}}, {Kent}, {Knapp}, {Lin}, {Loh}, {Lupton},
  {Margon}, {McKay}, {Meiksin}, {Munn}, {Pope}, {Richmond}, {Schlegel},
  {Schneider}, {Shimasaku}, {Stoughton}, {Strauss}, {SubbaRao}, {Szalay},
  {Szapudi}, {Tucker}, {Yanny}, \& {York}}]{Eisenstein+2005BAO}
{Eisenstein}, D.~J., {Zehavi}, I., {Hogg}, D.~W., {et~al.} 2005, {Detection of
  the Baryon Acoustic Peak in the Large-Scale Correlation Function of SDSS
  Luminous Red Galaxies}, {\em The Astrophysical Journal}, 633, 560

\bibitem[{Euclid(ca.~300~BC)}]{Euclid300BC}
Euclid. ca.~300~BC, {\em Elements} (R. Fitzpatrick), tr.~2008 R.~Fitzpatrick

\bibitem[{{Fry} \& {Seldner}(1982)}]{FrySeldner1982}
{Fry}, J.~N., \& {Seldner}, M. 1982, {Transform analysis of the high-resolution
  Shane-Wirtanen Catalog - The power spectrum and the bispectrum}, {\em The
  Astrophysical Journal}, 259, 474

\bibitem[{L\"ammel(2008)}]{Lammel2008}
L\"ammel, R. 2008, Google’s MapReduce programming model---Revisited, {\em
  Science of Computer Programming}, 70, 1

\bibitem[{Landau \& Lifshitz(1971)}]{LandauLifshitzFields}
Landau, L.~D., \& Lifshitz, E.~M. 1971, {\em The {{Classical Theory}} of
  {{Fields}}} (Pergamon Press)

\bibitem[{{Patterson}(1934)}]{Patterson1934}
{Patterson}, A.~L. 1934, {A Fourier Series Method for the Determination of the
  Components of Interatomic Distances in Crystals}, {\em Physical Review}, 46,
  372

\bibitem[{{Peebles}(1973)}]{Peebles1973}
{Peebles}, P.~J.~E. 1973, {Statistical Analysis of Catalogs of Extragalactic
  Objects. I. Theory}, {\em The Astrophysical Journal}, 185, 413

\bibitem[{{Peebles}(1980)}]{Peebles1980book}
{Peebles}, P.~J.~E. 1980, {\em {The Large-Scale Structure of the Universe}}
  (Princeton University Press)

\bibitem[{{Peebles} \& {Groth}(1975)}]{PeeblesGroth1975}
{Peebles}, P.~J.~E., \& {Groth}, E.~J. 1975, {Statistical Analysis of Catalogs
  of Extragalactic Objects. V. Three-point correlation function for the galaxy
  distribution in the Zwicky catalog.}, {\em The Astrophysical Journal}, 196, 1

\bibitem[{{Philcox} {et~al.}(2022){Philcox}, {Slepian}, {Hou}, {Warner},
  {Cahn}, \& {Eisenstein}}]{Philcox+2022ENCORE}
{Philcox}, O. H.~E., {Slepian}, Z., {Hou}, J., {et~al.} 2022, {ENCORE: an
  O(N$_{g}$$^{2}$) estimator for galaxy N-point correlation functions}, {\em
  Monthly Notices of the Royal Astronomical Society}, 509, 2457

\bibitem[{{Planck Collaboration} {et~al.}(2020{\natexlab{a}}){Planck
  Collaboration}, {Akrami}, {Arroja}, {Ashdown}, {Aumont}, {Baccigalupi},
  {Ballardini}, {Banday}, {Barreiro}, {Bartolo}, {Basak}, {Benabed}, {Bernard},
  {Bersanelli}, {Bielewicz}, {Bond}, {Borrill}, {Bouchet}, {Bucher},
  {Burigana}, {Butler}, {Calabrese}, {Cardoso}, {Casaponsa}, {Challinor},
  {Chiang}, {Colombo}, {Combet}, {Crill}, {Cuttaia}, {de Bernardis}, {de Rosa},
  {de Zotti}, {Delabrouille}, {Delouis}, {Di Valentino}, {Diego}, {Dor{\'e}},
  {Douspis}, {Ducout}, {Dupac}, {Dusini}, {Efstathiou}, {Elsner}, {En{\ss}lin},
  {Eriksen}, {Fantaye}, {Fergusson}, {Fernandez-Cobos}, {Finelli}, {Frailis},
  {Fraisse}, {Franceschi}, {Frolov}, {Galeotta}, {Galli}, {Ganga},
  {G{\'e}nova-Santos}, {Gerbino}, {Gonz{\'a}lez-Nuevo}, {G{\'o}rski},
  {Gratton}, {Gruppuso}, {Gudmundsson}, {Hamann}, {Handley}, {Hansen},
  {Herranz}, {Hivon}, {Huang}, {Jaffe}, {Jones}, {Jung}, {Keih{\"a}nen},
  {Keskitalo}, {Kiiveri}, {Kim}, {Krachmalnicoff}, {Kunz}, {Kurki-Suonio},
  {Lamarre}, {Lasenby}, {Lattanzi}, {Lawrence}, {Le Jeune}, {Levrier}, {Lewis},
  {Liguori}, {Lilje}, {Lindholm}, {L{\'o}pez-Caniego}, {Ma},
  {Mac{\'\i}as-P{\'e}rez}, {Maggio}, {Maino}, {Mandolesi}, {Marcos-Caballero},
  {Maris}, {Martin}, {Mart{\'\i}nez-Gonz{\'a}lez}, {Matarrese}, {Mauri},
  {McEwen}, {Meerburg}, {Meinhold}, {Melchiorri}, {Mennella}, {Migliaccio},
  {Miville-Desch{\^e}nes}, {Molinari}, {Moneti}, {Montier}, {Morgante}, {Moss},
  {M{\"u}nchmeyer}, {Natoli}, {Oppizzi}, {Pagano}, {Paoletti}, {Partridge},
  {Patanchon}, {Perrotta}, {Pettorino}, {Piacentini}, {Polenta}, {Puget},
  {Rachen}, {Racine}, {Reinecke}, {Remazeilles}, {Renzi}, {Rocha},
  {Rubi{\~n}o-Mart{\'\i}n}, {Ruiz-Granados}, {Salvati}, {Savelainen}, {Scott},
  {Shellard}, {Shiraishi}, {Sirignano}, {Sirri}, {Smith}, {Spencer}, {Stanco},
  {Sunyaev}, {Suur-Uski}, {Tauber}, {Tavagnacco}, {Tenti}, {Toffolatti},
  {Tomasi}, {Trombetti}, {Valiviita}, {Van Tent}, {Vielva}, {Villa},
  {Vittorio}, {Wandelt}, {Wehus}, {Zacchei}, \& {Zonca}}]{Planck18PNG}
{Planck Collaboration}, {Akrami}, Y., {Arroja}, F., {et~al.}
  2020{\natexlab{a}}, {Planck 2018 results. IX. Constraints on primordial
  non-Gaussianity}, {\em Astronomy \& Astrophysics}, 641, A9

\bibitem[{{Planck Collaboration} {et~al.}(2020{\natexlab{b}}){Planck
  Collaboration}, {Akrami}, {Arroja}, {Ashdown}, {Aumont}, {Baccigalupi},
  {Ballardini}, {Banday}, {Barreiro}, {Bartolo}, {Basak}, {Benabed}, {Bernard},
  {Bersanelli}, {Bielewicz}, {Bock}, {Bond}, {Borrill}, {Bouchet}, {Boulanger},
  {Bucher}, {Burigana}, {Butler}, {Calabrese}, {Cardoso}, {Carron},
  {Challinor}, {Chiang}, {Colombo}, {Combet}, {Contreras}, {Crill}, {Cuttaia},
  {de Bernardis}, {de Zotti}, {Delabrouille}, {Delouis}, {Di Valentino},
  {Diego}, {Donzelli}, {Dor{\'e}}, {Douspis}, {Ducout}, {Dupac}, {Dusini},
  {Efstathiou}, {Elsner}, {En{\ss}lin}, {Eriksen}, {Fantaye}, {Fergusson},
  {Fernandez-Cobos}, {Finelli}, {Forastieri}, {Frailis}, {Franceschi},
  {Frolov}, {Galeotta}, {Galli}, {Ganga}, {Gauthier}, {G{\'e}nova-Santos},
  {Gerbino}, {Ghosh}, {Gonz{\'a}lez-Nuevo}, {G{\'o}rski}, {Gratton},
  {Gruppuso}, {Gudmundsson}, {Hamann}, {Handley}, {Hansen}, {Herranz}, {Hivon},
  {Hooper}, {Huang}, {Jaffe}, {Jones}, {Keih{\"a}nen}, {Keskitalo}, {Kiiveri},
  {Kim}, {Kisner}, {Krachmalnicoff}, {Kunz}, {Kurki-Suonio}, {Lagache},
  {Lamarre}, {Lasenby}, {Lattanzi}, {Lawrence}, {Le Jeune}, {Lesgourgues},
  {Levrier}, {Lewis}, {Liguori}, {Lilje}, {Lindholm}, {L{\'o}pez-Caniego},
  {Lubin}, {Ma}, {Mac{\'\i}as-P{\'e}rez}, {Maggio}, {Maino}, {Mandolesi},
  {Mangilli}, {Marcos-Caballero}, {Maris}, {Martin},
  {Mart{\'\i}nez-Gonz{\'a}lez}, {Matarrese}, {Mauri}, {McEwen}, {Meerburg},
  {Meinhold}, {Melchiorri}, {Mennella}, {Migliaccio}, {Mitra},
  {Miville-Desch{\^e}nes}, {Molinari}, {Moneti}, {Montier}, {Morgante}, {Moss},
  {M{\"u}nchmeyer}, {Natoli}, {N{\o}rgaard-Nielsen}, {Pagano}, {Paoletti},
  {Partridge}, {Patanchon}, {Peiris}, {Perrotta}, {Pettorino}, {Piacentini},
  {Polastri}, {Polenta}, {Puget}, {Rachen}, {Reinecke}, {Remazeilles}, {Renzi},
  {Rocha}, {Rosset}, {Roudier}, {Rubi{\~n}o-Mart{\'\i}n}, {Ruiz-Granados},
  {Salvati}, {Sandri}, {Savelainen}, {Scott}, {Shellard}, {Shiraishi},
  {Sirignano}, {Sirri}, {Spencer}, {Sunyaev}, {Suur-Uski}, {Tauber},
  {Tavagnacco}, {Tenti}, {Toffolatti}, {Tomasi}, {Trombetti}, {Valiviita}, {Van
  Tent}, {Vielva}, {Villa}, {Vittorio}, {Wandelt}, {Wehus}, {White}, {Zacchei},
  {Zibin}, \& {Zonca}}]{Planck18Inflation}
{Planck Collaboration}, {Akrami}, Y., {Arroja}, F., {et~al.}
  2020{\natexlab{b}}, {Planck 2018 results. X. Constraints on inflation}, {\em
  Astronomy \& Astrophysics}, 641, A10

\bibitem[{{Portillo} {et~al.}(2018){Portillo}, {Slepian}, {Burkhart},
  {Kahraman}, \& {Finkbeiner}}]{Portillo+2018}
{Portillo}, S. K.~N., {Slepian}, Z., {Burkhart}, B., {Kahraman}, S., \&
  {Finkbeiner}, D.~P. 2018, {Developing the 3-point Correlation Function for
  the Turbulent Interstellar Medium}, {\em The Astrophysical Journal}, 862, 119

\bibitem[{{Slepian} \& {Eisenstein}(2016)}]{SlepianEisenstein2016}
{Slepian}, Z., \& {Eisenstein}, D.~J. 2016, {Accelerating the two-point and
  three-point galaxy correlation functions using Fourier transforms}, {\em
  Monthly Notices of the Royal Astronomical Society}, 455, L31

\bibitem[{Su {et~al.}(2024)Su, Ahmed, Lu, Pan, Bo, \& Liu}]{Su2024Roformer}
Su, J., Ahmed, M., Lu, Y., {et~al.} 2024, Roformer: Enhanced transformer with
  rotary position embedding, {\em Neurocomputing}, 568, 127063

\bibitem[{Vaswani {et~al.}(2017)Vaswani, Shazeer, Parmar, Uszkoreit, Jones,
  Gomez, Kaiser, \& Polosukhin}]{Vaswani2017Attention}
Vaswani, A., Shazeer, N., Parmar, N., {et~al.} 2017, Attention is all you need,
  {\em Advances in Neural Information Processing Systems}, 30

\bibitem[{Waring(1782)}]{Waring1782}
Waring, E. 1782, {\em Meditationes Algebraicae} (American Mathematical
  Society), tr.~1991 D.~Weeks

\bibitem[{Weyl(1939)}]{Weyl1939}
Weyl, H. 1939, {\em The Classical Groups: Their Invariants and Representations}
  (Princeton University Press)

\bibitem[{Zaheer {et~al.}(2017)Zaheer, Kottur, Ravanbakhsh, Poczos,
  Salakhutdinov, \& Smola}]{Zaheer+17deepsets}
Zaheer, M., Kottur, S., Ravanbakhsh, S., {et~al.} 2017, Deep Sets, in {\em
  Advances in Neural Information Processing Systems}, ed. I.~Guyon {et~al.},
  Vol.~30 (Curran Associates, Inc.)

\end{thebibliography}

\secbreak
\appendix
\subsection{Appendix}
\label{sec:appendix}

\paragraph{Additional formulae} We noted that symbolic regression revealed several equivalent but more complicated triangle area formulae which are polynomials in the multi-power sums. Below is a list of these.

For brevity, we define
\begin{equation}
    F_{mn} \equiv \sum_{i=1}^{3} \, x_i^m \, y_i^n.
\end{equation}
For reference, our result \eqref{eq:result} is
\begin{equation}
4\,\Delta^2 =
+3\,F_{20}\,F_{02}
-3\,F_{11}^2
+2\,F_{11}\,F_{10}\,F_{01}
-1\,F_{20}\,F_{01}^2
-1\,F_{02}\,F_{10}^2~.
\end{equation}
The additional formulae are
\begin{align}
4\,\Delta^2 =
&
+6\,F_{22}
-4\,F_{21}\,F_{01}
-4\,F_{12}\,F_{10}
+2\,F_{20}\,F_{02}
\nonumber\\ &
+6\,F_{11}\,F_{10}\,F_{01}
-5\,F_{11}^2
-1\,F_{10}^2\,F_{01}^2~,
\\
4\,\Delta^2 =
&
+18\,F_{22}
-12\,F_{21}\,F_{01}
-12\,F_{12}\,F_{10}
\nonumber\\ &
+2\,F_{20}\,F_{01}^2
+2\,F_{02}\,F_{10}^2
+14\,F_{11}\,F_{10}\,F_{01}
-9\,F_{11}^2
-3\,F_{10}^2\,F_{01}^2~,
\\
8\,\Delta^2 =
&
-6\,F_{22}
+4\,F_{21}\,F_{01}
+4\,F_{12}\,F_{10}
+7\,F_{20}\,F_{02}
\nonumber\\ &
-3\,F_{20}\,F_{01}^2
-3\,F_{02}\,F_{10}^2
-4\,F_{11}^2
+1\,F_{10}^2\,F_{01}^2~,
\\
8\,\Delta^2 =
&
-18\,F_{22}
+12\,F_{21}\,F_{01}
+12\,F_{12}\,F_{10}
+9\,F_{20}\,F_{02}
\nonumber\\ &
-5\,F_{20}\,F_{01}^2
-5\,F_{02}\,F_{10}^2
-8\,F_{11}\,F_{10}\,F_{01}
+3\,F_{10}^2\,F_{01}^2~,
\end{align}
and, our personal favorite,
\begin{align}
368\,\Delta^2 =
&
-54\,F_{22}
+36\,F_{21}\,F_{01}
+36\,F_{12}\,F_{10}
\nonumber\\ &
+285\,F_{20}\,F_{02}
-101\,F_{20}\,F_{01}^2
-101\,F_{02}\,F_{10}^2
\nonumber\\ &
-258\,F_{11}^2
+148\,F_{11}\,F_{10}\,F_{01}
+9\,F_{10}^2\,F_{01}^2~.
\end{align}
We note that 368 is divisible by 23.

\paragraph{Proof that the area sums in quadrature} We prove equation \eqref{eq:quadrature}, that the area of a triangle in $d$ dimensions is the sum in quadrature of the areas of the triangles made by all pairs of mutually orthogonal projections of the points.

From the three position vectors $\vec{r}_1, \vec{r}_2, \vec{r}_3 \in \mathbb{R}^d$ (for $d \geq 2$), we consider the two displacement vectors $\vec{a} \equiv \vec{r}_2 - \vec{r}_1$ and $\vec{b} \equiv \vec{r}_3 - \vec{r}_1$.
Given these, we can write the two-dimensional triangle sub-area $\Delta_{jk}$ by first defining the projected two-dimensional displacement vectors $\vec{a}_{jk} = \big[ a_j, a_k \big]$, $\vec{b}_{jk} = \big[ b_j, b_k \big]$. Then,
\begin{equation}
    \Delta_{jk}
    = \frac{1}{2} \, \abs{ \vec{a}_{jk} \wedge \vec{b}_{jk} }
    = \frac{1}{2}
        \, \abs{a_j \, b_k - b_j \, a_k}~,
\end{equation}
using formula \eqref{eq:cross}.

Next, consider formula \eqref{eq:cross} with the full $d$-dimensional vectors:
\begin{align}
    \Delta^2
    &= \frac{1}{4}\, \abs{\vec{a} \wedge \vec{b}}^2 \nonumber\\
    &= \frac{1}{4}\, \abs{\vec{a}}^2 \, \abs{\vec{b}}^2 - \frac{1}{4}\, \abs{\vec{a} \cdot \vec{b}}^2 \nonumber\\
    &= \frac{1}{4} \sum_{1 \leq j, k \leq d}
        ( a_j \, a_j \, b_k \, b_k - a_j \, b_j \, a_k \, b_k ) \nonumber\\
    &= \frac{1}{4} \sum_{1 \leq j < k \leq d}
        ( a_j \, b_k - b_j \, a_k )^2 \nonumber\\
    &= \sum_{1 \leq j < k \leq d} \Delta_{jk}^2~.
\end{align}

\paragraph{Volume of the $d$-simplex} Consider a $d$-simplex in $d$-dimensional space. Let its vertices be labeled $\{ \vec{r}_1, \dots, \vec{r}_{d+1} \}$. The volume of the $d$-simplex is known to be
\begin{equation}
    V = \frac{1}{d!} \, \abs{\det \mathsf{A}},
    \quad \mathsf{A} \equiv \left[ \begin{array}{cc}
        \vec{r}_1 & 1 \\
        \vdots & \vdots \\
        \vec{r}_{d+1} & 1 \\
    \end{array} \right]~,
\end{equation}
where $\mathsf{A}$ is a $(d+1) \times (d+1)$ matrix whose rows are the vertices, appended by a single 1.

Squaring both sides, we find
\begin{equation}
    V^2
    = \frac{1}{(d!)^2} \, (\det \mathsf{A})^2
    = \frac{1}{(d!)^2} \, \det \mathsf{A}^\top \mathsf{A},
\end{equation}
and $\mathsf{A}^\top \mathsf{A}$ matches $\mathsf{M}$ as defined in equation \eqref{eq:dsimplex}.

\paragraph{Deep Sets form for the energy} Consider a system of $N$ point particles with masses $m_i$ and positions $\vec{r}_i$.
The energy density $\mathcal{E} = \d U / \d^3 x$ is given by
\begin{equation}
    \mathcal{E} = \frac{\varrho \, \Phi,}{2}~,
\end{equation}
where $\varrho$ is the mass density and $\Phi$ is the gravitational potential \citep{LandauLifshitzFields}.
These are related by the Poisson equation
\begin{equation}
    \nabla^2 \Phi = 4 \pi G \, \varrho~,
\end{equation}
which implies the following relationship between their Fourier transforms
\begin{equation}
    -k^2 \, \mathcal{F}\{\Phi\} = 4 \pi G \, \mathcal{F}\{\varrho\}~,
\end{equation}
with $k^2 \equiv \abs{\vec{k}}^2$.
Using the convolution theorem, we then have
\begin{align}
    \mathcal{F}\{\mathcal{E}\}
    &= \frac{1}{2} \, \mathcal{F}\{\varrho \, \Phi\} \nonumber\\
    &= \frac{1}{2 \, (2\pi)^3} \, \mathcal{F}\{\varrho\} \ast \mathcal{F}\{\Phi\} \nonumber\\
    &= -\frac{G}{4\pi^2} \, \mathcal{F}\{\varrho\} \ast \left( k^{-2} \, \mathcal{F}\{\varrho\} \right)~.
    \label{eq:ConvolutionThm}
\end{align}
This form of the convolution theorem is proven below.
Thus, the total energy is given in terms of the Fourier transform of the density field by
\begin{equation}
    U 
    = -\frac{G}{4 \pi^2} \,
        \int \d^3 r \,
        \mathcal{F}^{-1} \! \left\{
            \mathcal{F}\{\varrho\} \ast \left(
            k^{-2} \, \mathcal{F}\{\varrho\} \right)
        \right\}~.
\end{equation}

\paragraph{Proof of the convolution theorem} We prove the claim (equation \eqref{eq:ConvolutionThm}) that
\begin{equation}
    \mathcal{F}\{\varrho \, \Phi\}
    = (2\pi)^{-3} \, \mathcal{F}\{\varrho\} \ast \mathcal{F}\{\Phi\} ~.
\end{equation}
Note that we use the conventions:
\begin{align}
    \tilde{f}(\vec{k})
    = \mathcal{F}\{f(\vec{x})\}
    &= \int \d^3 x
        \, e^{-i \, \vec{k} \cdot \vec{x}}
        \, f(\vec{x})~, \\
    f(\vec{x})
    = \mathcal{F}^{-1}\{\tilde{f}(\vec{k})\}
    &= \int \frac{\d^3 k}{(2\pi)^3}
        \, e^{+i \, \vec{k} \cdot \vec{x}}
        \, \tilde{f}(\vec{k})~.
\end{align}
Consider $\mathcal{F}^{-1}\{\tilde{\varrho} \ast \tilde{\Phi}\}$. 
\begin{align}
    \mathcal{F}^{-1}\{\tilde{\varrho} \ast \tilde{\Phi}\}(\vec{x})
    &= \int_{\mathbb{R}^3} \frac{\d^3 k}{(2\pi)^3}
        \, e^{i \, \vec{k} \cdot \vec{x}} 
        \, \big( \tilde{\varrho} \ast \tilde{\Phi} \big) (\vec{k}) 
        \nonumber\\
    &= \int_{\mathbb{R}^3} \frac{\d^3 k}{(2\pi)^3}
        \, e^{i \, \vec{k} \cdot \vec{x}} 
        \, \int_{\mathbb{R}^3} \d^3 \vec{k}' 
        \, \tilde{\varrho}(\vec{k}') 
        \, \tilde{\Phi}(\vec{k} - \vec{k}')
        \nonumber\\
    &= \int_{\mathbb{R}^3} \d^3 \vec{k}'
        \, \tilde{\varrho}(\vec{k}') 
        \, \int_{\mathbb{R}^3} \frac{\d^3 k}{(2\pi)^3}
        \, e^{i \, \vec{k} \cdot \vec{x}} 
        \, \tilde{\Phi}(\vec{k} - \vec{k}')
        \nonumber\\
    &= \int_{\mathbb{R}^3} \d^3 \vec{k}'
        \, \tilde{\varrho}(\vec{k}') 
        \, e^{i \, \vec{k}' \cdot \vec{x}}
        \, \Phi(\vec{x})
        \nonumber\\
    &= (2\pi)^3
        \, \varrho(\vec{x}) 
        \, \Phi(\vec{x})~.
\end{align}
Taking the Fourier transform of both sides gives the desired result.

\paragraph{Fourier-space triangle area formula}
Suppose our triangle is specified by three vertices $\vec{r}_1,\vec{r}_2,\vec{r}_3 \in \mathbb{R}^d$. We represent this distribution by the density field
\begin{equation}
    \varrho(\vec{r})
    = \sum_i \, \delta(\vec{r} - \vec{r}_i)~.
\end{equation}
The Fourier transform of this field is
\begin{equation}
    \tilde{\varrho}(\vec{k})
    =
    e^{-i \, \vec{k} \cdot \vec{r}_1} +
    e^{-i \, \vec{k} \cdot \vec{r}_2} +
    e^{-i \, \vec{k} \cdot \vec{r}_3}~.
\end{equation}
Multiplying by its conjugate to find the magnitude squared, we find
\begin{equation}
    \abs{\tilde{\varrho}(\vec{k})}^2
    =
    3 +
    e^{-i \, \vec{k} \cdot \vec{r}_{12}} +
    e^{+i \, \vec{k} \cdot \vec{r}_{12}} +
    e^{-i \, \vec{k} \cdot \vec{r}_{13}} +
    e^{+i \, \vec{k} \cdot \vec{r}_{13}} +
    e^{-i \, \vec{k} \cdot \vec{r}_{23}} +
    e^{+i \, \vec{k} \cdot \vec{r}_{23}}~,
\end{equation}
defining $\vec{r}_{ij} \equiv \vec{r}_i - \vec{r}_j$.
The inverse Fourier transform of this is
\begin{align}
    \mathcal{P}(\vec{r})
    &= (2\pi)^d \, \mathcal{F}^{-1} \! \left\{ \abs{\tilde{\varrho}}^2 \right\}
    \nonumber\\ 
    &= 3 \,\delta(\vec{r})
    + \delta(\vec{r} - \vec{r}_{12})
    + \delta(\vec{r} + \vec{r}_{12})
    \nonumber\\
    &\quad + \delta(\vec{r} - \vec{r}_{13})
    + \delta(\vec{r} + \vec{r}_{13})
    + \delta(\vec{r} - \vec{r}_{23})
    + \delta(\vec{r} + \vec{r}_{23})~.
\end{align}
This is known as the Patterson function in x-ray crystallography.

By multiplying the Patterson function $\mathcal{P}$ by a function of $r \equiv \abs{\vec{r}}$ and integrating over all space, we can compute sums of functions of the side lengths.
\begin{align}
    \int_{\mathbb{R}^d} \d^d r \, \mathcal{P}(\vec{r}) \, f(\abs{\vec{r}})
    &= 3 \, f(0) +
        2 \, f(r_{12}) +
        2 \, f(r_{13}) +
        2 \, f(r_{23}) \nonumber\\
    &= 3 \, f(0) +
        2 \, f(a) +
        2 \, f(b) +
        2 \, f(c)~.
\end{align}
In particular, we can compute the $m^{\rm th}$ power sum of the side lengths
\begin{equation}
    P_m
    = \int_{\mathbb{R}^d} \d^d r \, \mathcal{P}(\vec{r}) \, \frac{\abs{\vec{r}}^m}{2}
    = a^m + b^m + c^m~,
\end{equation}
which allows us to write the triangle area as
\begin{equation}
    \Delta
    = \frac{1}{4} \, \sqrt{\left(a^2+b^2+c^2\right)^2 - 2\, \left(a^4+b^4+c^4\right)}
    = \frac{1}{4} \, \sqrt{P_2^2 - 2\, P_4}~.
\end{equation}

\end{document}